\documentclass[a4paper,12pt]{article}
\usepackage[active]{srcltx}
\usepackage{hyperref}
\usepackage{amsmath,amssymb,epsf,epic}
\usepackage{times}
\usepackage{mathrsfs}
\usepackage{latexsym,amsopn}
\usepackage{amsfonts,amsbsy,amscd,stmaryrd}
\usepackage{theorem,enumerate}
\usepackage{paralist}
\usepackage{geometry}
\numberwithin{equation}{section}

\newcommand{\tr}{{\rm Tr}}

\newcommand{\bra}{\langle} 
\newcommand{\ket}{\rangle}

\newcommand{\gs}{|0\rangle}
\newcommand{\es}{|1\rangle}

\newcommand{\vd}{v_{\mathrm d}}

\newcommand{\e}{{\rm e}}

\newcommand{\p}{{\rm p}}
\renewcommand{\a}{{\rm a}}
\let\II\i
\renewcommand{\i}{{\rm i}}
\renewcommand{\d}{{\rm d}}

\renewcommand{\sp}{{\rm sp}}

\newcommand{\tH}{\widetilde H}

\newcommand{\grintl}{[\kern-.18em [}
\newcommand{\grintr}{]\kern-.18em ]}

\def\env{{\rm env}}

\newcounter{resultcounter}[section]

\newtheorem{theorem}[resultcounter]{Theorem}
\newtheorem{lemma}[resultcounter]{Lemma}
\newtheorem{proposition}[resultcounter]{Proposition}
\newtheorem{corollary}[resultcounter]{Corollary}
\newtheorem{definition}[resultcounter]{Definition}
\newtheorem{remark}[resultcounter]{Remark}

\newcommand\bep{\begin{proposition}}
\newcommand\eep{\end{proposition}}
\newcommand\ber{\begin{remark}}
\newcommand\eer{\end{remark}}
\newcommand\bel{\begin{lemma}}
\newcommand\eel{\end{lemma}}
\newcommand\bet{\begin{theorem}}
\newcommand\eet{\end{theorem}}
\newcommand\bex{\begin{example}}
\newcommand\eex{\end{example}}
\newcommand\bed{\begin{definition}}
\newcommand\eed{\end{definition}}
\newcommand\bea{\begin{assumption}}
\newcommand\eea{\end{assumption}}
\newcommand\bec{\begin{corollary}}
\newcommand\eec{\end{corollary}}
\newcommand{\beq}{\begin{equation}}
\newcommand{\eeq}{\end{equation}}
\newcommand{\beqa}{\begin{eqnarray}}
\newcommand{\eeqa}{\end{eqnarray}}

\def\cA{{\cal A}} \def\cB{{\cal B}} \def\cC{{\cal C}}
\def\cD{{\cal D}}  
 \def\cH{{\cal H}} 
\def\cJ{{\cal J}} \def\cK{{\cal K}} \def\cL{{\cal L}}

  \def\cU{{\cal U}}

\newcommand{\R}{{\mathbb R}}
\newcommand{\N}{{\mathbb N}}
\newcommand{\Q}{{\mathbb Q}}
\newcommand{\C}{{\mathbb C}}
\newcommand{\Z}{{\mathbb Z}}
\newcommand{\E}{{\mathbb E}}
\renewcommand{\P}{{\mathbb P}}

\def\qed{\hfill $\Box$\medskip}

\begin{document}
\title{Scattering induced current in a tight binding band}

\author{L. Bruneau\thanks{laurent.bruneau@u-cergy.fr}\\
D\'ept. de Math\'ematiques and UMR 8088, CNRS\\
Universit\'e de Cergy-Pontoise\thanks{95000 Cergy-Pontoise, France}\\
and\\
S. {\ De Bi\`{e}vre}\thanks{Stephan.De-Bievre@math.univ-lille1.fr} \\
Lab. P. Painlev\'e, CNRS, UMR 8524 and UFR de Math\'ematiques \\
Universit\'e Lille 1, Sciences et Technologies\thanks{F-59655 V'd'Ascq Cedex, France}\\
\'Equipe-Projet SIMPAF,
Centre de Recherche INRIA Futurs\thanks{Parc Scient. de la Haute Borne, 40, avenue Halley B.P. 70478,
F-59658 V'd'Ascq cedex, France}\\
and\\
C.-A. Pillet\thanks{pillet@univ-tln.fr}\\
Centre de Physique Th\'eorique\footnote{%
UMR 6207:
Univ. de Provence,
Univ. de la M\'editerran\'ee, 
Univ. de Toulon et CNRS, FRUMAM}\\
Universit\'e du Sud Toulon-Var\thanks{B.P. 20132,
83957 La Garde Cedex, France}
}
\date{\today}
\maketitle

\bigskip
{\bf\noindent Abstract.}
In the single band tight-binding approximation, we consider the transport properties of an electron
subject to a homogeneous static electric field. We show that repeated interactions of the electron
with two-level systems in thermal equilibrium suppress the Bloch oscillations and induce a steady current, the
statistical properties of which we study.

\section{Introduction}
Let us consider the motion of an electron in the single band tight-binding approximation. It is well 
known that in the absence of a dc electric field, the electron moves ballistically whereas, when a dc
electric field is present, Bloch oscillations prevent a current from being set up in the system. 
It is furthermore expected that if the electron is in contact with a thermal environment, the resulting
scattering mechanisms will suppress the Bloch oscillations and lead to a steady current. This 
phenomenon can for example be described within a semi-classical picture of the motion combined 
with the relaxation-time approximation (see~\cite{am} for example). Alternatively, in open systems
theory, one describes the thermal environment and its coupling to the electron with a global 
Hamiltonian, and traces out the environment variables to obtain, under suitable additional conditions
such as weak coupling and appropriate scalings, effective dynamical equations for the electron alone.
In treatments of this type, the environment is often described by a set of oscillators (see for 
example~\cite{fz}) or more generally by a free Bose field. For a review of various approaches to
transport theory we refer to~\cite{bel}. There has recently been intensified interest in obtaining
rigorous results along those lines~\cite{dfp, df, cdm}.

We present here a simple, explicitly solvable, fully quantum mechanical and fully Hamiltonian model 
of such a particle-environment system within the repeated interaction scheme, in which the 
environment is described by a chain of two-level atoms. We show that a dc current is created due 
to the interaction of the particle with its environment. In addition to drifting in the direction of the 
applied field, the electron diffuses around its mean position. We give a full analysis of the probability distribution for the position of the particle in the large time regime (Theorem \ref{thm:main}). We then use a repeated measurement scheme to describe the increments of the position and energy observables between time $0$ and time $t$ (Theorems \ref{thm:FCSx} and \ref{thm:FCSenergy}).

The rest of the paper is organized as follows. In Section~\ref{sec:model} we give a detailed 
description of the model under consideration. Our main results are stated precisely in 
Section~\ref{sec:results}. We describe the effect on the particle of the interaction with a single atom in Section~\ref{s:oneatom}, and the main properties of the repeated interaction dynamics in Section~\ref{s:proofs}. The proofs of the main theorems are provided in Sections~\ref{s:proofthm1} and~\ref{s:FCS}.

\section{Description of the model}
\label{sec:model}

We consider a spinless particle on the one-dimensional lattice $\Z$ and submitted to a constant
external force $F\geq 0$. The quantum Hilbert space and Hamiltonian of the particle are
\begin{equation}
\label{def:particle}
\cH_{\mathrm p}=\ell^2(\Z), \qquad H_{\mathrm p}=-\Delta -FX,
\end{equation}
where $\Delta$ is the usual discrete nearest neighbor Laplacian and $X$ the lattice position operator
$$
-\Delta =\sum_{x\in\Z} \bigl(2\,|x\rangle\langle x|-|x\!+\!1\rangle\langle x| - | x\rangle\langle x\!+\!1|\bigr),
\quad
X=\sum_{x\in\Z}x \,|x\rangle\langle x|.
$$
Identifying $\cH_{\mathrm p}$ with  $L^2(\mathbb T^1,\d\xi)$ via the discrete Fourier transform,
we have
$$
-\Delta=2(1-\cos\xi),\qquad X=\i\partial_\xi.
$$
Here $\mathbb T  ^1\simeq [0,2\pi[$ is the first Brillouin zone and $\xi$  the crystal momentum.
Defining the translation operator
$$
T=\sum_{x\in\Z} |x\!+\!1\rangle\langle x|=\e^{-\i\xi},
$$ 
we can write $H_{\mathrm p}=2-T-T^*-FX$. We note for later reference that
\begin{equation}
\label{PartCR}
[X,T]=T,\quad [X,T^*]=-T^*,\quad [H_\p, T]=-FT, \quad [H_\p,T^*]=FT^*.
\end{equation}
When $F=0$, $H_\p$ has a single band of absolutely continuous spectrum,
$\sp(H_\p)=[0,4]$, and the motion of the particle is described by
\begin{eqnarray*}
T(t)&=&\e^{\i tH_\p}T\e^{-\i tH_\p}=T,\\
X(t)&=&\e^{\i tH_\p}X\e^{-\i tH_\p}=X+\i(T-T^\ast)t,
\end{eqnarray*}
showing its ballistic nature.

When $F\not=0$, we set $G=2F^{-1}\sin\xi=\i F^{-1}(T-T^\ast)$. The commutation relations
(\ref{PartCR}) yield
\begin{equation}
\label{HpConjug}
H_\p=\e^{\i G}(2-FX)\e^{-\i G},
\end{equation}
from which it follows that $H_\p$
has discrete spectrum, $\sp(H_\p)=2-F\Z$. This is the well-known Wannier-Stark ladder.
In the momentum representation, the normalized eigenvector $\psi_k$ to the eigenvalue
$E_k=2-Fk$ is given by
\begin{equation}\label{eq:psik}
\widehat\psi_k(\xi)=\frac{1}{\sqrt{2\pi}}\,\e^{\i(2F^{-1}\sin(\xi)-k\xi)}.
\end{equation}
In the position representation, we therefore have
$$
\psi_k(x)=\int_0^{2\pi} \e^{\i(2F^{-1}\sin(\xi)-(k-x)\xi)}\,\frac{\d\xi}{2\pi}=J_{k-x}\left(\frac{2}{F} \right),
$$
where the $J_\nu$ are Bessel functions. From their asymptotic behavior for large $\nu$
(see e.g. Formula (10.19.1) in \cite{OLBC}) we infer that
$$
\psi_k(x)\sim\frac1{\sqrt{2\pi|k-x|}}\left(\frac\e{F|k-x|}\right)^{|k-x|}
\quad\text{for }|k-x|\to\infty,
$$
which shows that  $\psi_k(x)$ is sharply localized around $x=k$. The motion of the particle,
described by
\begin{equation}
\label{eq:xfreeevol}
\left.
\begin{array}{rcl}
\displaystyle T(t)&=&\e^{\i tH_\p}T\e^{-\i tH_\p}=\e^{-\i tF}T,\\[10pt]
X(t)&=&\displaystyle\e^{\i t H_\p}X\e^{-\i t H_\p}
=X+\frac{4}{F}\sin\left(\frac{Ft}{2} \right)\sin\left(\xi+\frac{Ft}{2}\right),
\end{array}
\right\}
\end{equation}
is now confined by Bloch oscillations.

\medskip

In what follows, we let the particle interact with one or many 2-level atoms, each of which has 
a quantum Hilbert space $\cH_\a=\C^2$ which we identify with $\Gamma_-(\C)$, the fermionic
Fock space over $\C$. The atomic Hamiltonian is given by
$$
H_\a=\d\Gamma(E)=Eb^*b,
$$
where $E\geq 0$ is the Bohr frequency of the atom and $b^\ast$, $b$ are the usual Fermi creation 
and annihilation operators. The later satisfy the canonical anti-commutation relations
\begin{equation}
\label{CAR}
b^*b + bb^* = I,
\qquad
b^2=b^{\ast2}=0.
\end{equation}
The initial state of the two-level atoms will be their equilibrium state at inverse temperature $\beta$
described by the density matrix
\begin{equation}
\label{def:thermalstate}
\rho_\beta=Z_\beta^{-1}\e^{-\beta H_\a}, \quad  Z_\beta= \tr (\e^{-\beta H_\a})=1+\e^{-\beta E}.
 \end{equation}
The interaction between the particle and the two-level atom is chosen so that its effect is to give 
a right or left kick to the particle, depending on whether the atom is in its ground state or in its
excited state. More precisely, we set
\begin{equation}
\label{def:interaction}
V=\sum_{x\in\Z}\bigl( |x\!+\!1\ket\bra x| \otimes b^* + |x\ket\bra x\!+\!1| \otimes b\bigr)
=Tb^*+T^*b.
\end{equation}
To understand this interaction, note that when $F>0$, the translation operator $T$ can be
thought of as a lowering operator for the particle. Indeed, from (\ref{eq:psik}) one finds
\begin{equation}
\label{eq:Tpsi}
T\psi_k=\psi_{k+1}
\end{equation}
Similarly, $T^*$ acts as a raising operator. As a result, $V$ describes an exchange of energy
between the two-level system and the particle. We point out that this interaction is very similar
to the one which appears in the Jaynes-Cummings Hamiltonian where a two level atom interacts
with one mode of the electromagnetic field of a cavity (through its electric dipole moment and
in the rotating wave approximation), see e.g. \cite{CDG,Du}. Thermalization of the field
through repeated interaction with two-level atoms was proven for the Jaynes-Cummings
Hamiltonian in~\cite{BP}. The model treated here is very similar to the one studied in~\cite{BP},
except that the spectrum of $H_\p$, contrary to the spectrum of the mode of the
electromagnetic field, is not  bounded from below. As a result, the system we treat here has
no invariant state, as we shall see below.

The full Hamiltonian of the particle interacting with a single two-level system acts on
the Hilbert space $\cH_\p\otimes \cH_\a$ and is given by
\begin{equation}
\label{eq:honeatom}
H=H_\p+H_\a+\lambda V,
\end{equation}
where $\lambda\in\R$ is a coupling constant. As the more explicit formula
$$
H=2-T(1-\lambda b^*)-T^*(1-\lambda b)-FX+Eb^*b,
$$
shows, one can also interpret the coupling to the two-level system as altering the hopping
matrix elements of the original Hamiltonian. The operator $H$ is easily diagonalized by noticing
that it commutes with the ``number operator''
\begin{equation}
\label{def:Nop}
N=-\e^{\i G}X\e^{-\i G}+b^\ast b=\frac{H_\p-2}F+\frac{H_\a}E,
\end{equation}
which has a two-dimensional eigenspace to each of its eigenvalue $n\in\Z$.
In particular, if $E=F$ then the energy $H_\p+H_\a$ is preserved by the full dynamics
(which will be computed in Section~\ref{s:oneatom}).


We now turn to the description of the repeated interaction dynamics (see e.g. \cite{AP,BJM,BP}).
We let the particle interact successively, each time during a fixed period $\tau>0$, with the
elements of a sequence of atoms, {\sl i.e.,} during the time interval $[(n-1)\tau,n\tau[$, the particle
interacts with the $n$-th atom and with none of the others. The Hilbert space of the atomic
reservoir is
$$
 \cH_\env=\bigotimes_{n=1}^M\cH_{\a,n},
$$
where $M$ is the number of atoms and each $\cH_{\a,n}$ is a copy of $\cH_\a$. 
The Hilbert space of the joint particle+reservoir system is
$
\cH=\cH_{\mathrm p}\otimes\cH_\env.
$
The full unitary evolution $U(t,s)$ of the system is thus described by the Schr\"odinger equation
\begin{equation}
\label{eq:unitevol}
i\partial_tU(t,s)=H(t)U(t,s),\qquad U(s,s)=I,\qquad t,s\in[0,M\tau],
\end{equation}
with time-dependent Hamiltonian 
\begin{equation}\label{eq:totalH}
H(t)=H_{\mathrm p}+ \sum_{n=1}^MH_{\a,n} +\lambda\sum_{n=1}^M\chi_n(t)(Tb^*_n+T^*b_n),
\end{equation}
where $\chi_n$ is the characteristic function of the interval $[(n-1)\tau,n\tau[$ and $H_{\a,n}$,
$b_n$, $b_n^*$ are the Hamiltonian, annihilation and creation operators of the $n$-th atom.
We will use the following notation
\begin{equation}
\label{eq:Han}
H_n= H_\p+H_{\a,n}+\lambda(Tb^*_n+T^*b_n),
\qquad
\tH_n= H_n+\sum_{{1\le k\le M}\atop{k\not=n}} H_{\a,k}.
\end{equation}
Note that $H(t)=\tH_n$ when $t\in [(n-1)\tau,n\tau[$. 

We denote by ${\cal B}^1(\cal H_\p)$ the Banach space of trace class operators on the Hilbert space 
$\cal H_\p$. Given any density matrix for the particle $\rho_{\mathrm p}\in{\cal B}^1({\cal H}_\p)$, 
$\rho_\p\ge0$, $\tr\rho_\p=1$, we set the initial state of the joint system to
$$
\rho_0=\rho_\p\otimes\rho_\beta^{\otimes M},\qquad
\rho_\beta^{\otimes M}=\bigotimes_{n=1}^M\rho_\beta.
$$
After $n\le M$ interactions, this state evolves into $\rho_{n\tau}=U(n\tau,0)\rho_0 U(n\tau, 0)^*$.
To obtain the density matrix $\rho_{\p,n\tau}$ of the particle after these $n$ interactions we 
take the partial trace over the environment,
\begin{equation}
\label{eq:rhoptn}
\rho_{\p,n\tau}=\tr_{\cH_\env}\rho_{n\tau}=\tr_{\cH_\env}
U(n\tau,0)(\rho_\p\otimes\rho_\beta^{\otimes M})U(n\tau, 0)^*.
\end{equation}
In fact, for reasons that will become clear later, we shall consider the more general linear operator 
defined on $\cB(\cH_\p)$ by
$$
D_{\alpha,n}(A)=\tr_{\cH_\env}\left(I\otimes\left[\rho_\beta^{\alpha}\right]^{\otimes M}\right)
U(n\tau,0)\left(A\otimes\left[\rho_\beta^{1-\alpha}\right]^{\otimes M}\right)
U(n\tau,0)^\ast,
$$
where $\alpha$ is an arbitrary real parameter. Using the cyclicity of the partial trace w.r.t. atomic
operators and the fact that $\rho_\beta$ is invariant under the free atomic dynamics we can
replace $U(n\tau,0)$ by $\e^{-\i\tau H_n}U((n-1)\tau,0)$ in the last formula. It then follows that
$$
D_{\alpha,n}(A)
=\tr_{\cH_\a}(I\otimes\rho_\beta^{\alpha})\e^{-\i\tau H}
(D_{\alpha,n-1}(A)\otimes\rho_\beta^{1-\alpha})\,\e^{\i\tau H}=\cL_\alpha(D_{\alpha,n-1}(A)),
$$
where
\begin{equation}
\label{def:Lalpha}
\cL_\alpha(A)=\tr_{\cH_\a}(I\otimes\rho_\beta^{\alpha})
\e^{-\i\tau H}(A\otimes\rho_\beta^{1-\alpha})\,\e^{\i\tau H}.
\end{equation}
We conclude that $D_{\alpha,n}(A)=\cL_\alpha^n(A)$ and hence
\begin{equation}
\label{LalphaMeans}
\tr_{\cH_\p}B\cL_\alpha^n(A)=
\tr(B\otimes\left[\rho_\beta^\alpha\right]^{\otimes M})U(n\tau,0)
(A\otimes\left[\rho_\beta^{1-\alpha}\right]^{\otimes M})U(n\tau,0)^\ast.
\end{equation}
The operator $\cL_\alpha$ will play a central role in our analysis. For later reference, we
describe its main properties in the following proposition.

\bep\label{prop:Lalpha} For any $\alpha\in\R$, $\cL_\alpha$ is a completely positive operator on $\cB^1(\cH_\p)$
with spectral radius
\begin{equation}\label{eq:thetadef}
\theta(\alpha)=(1-p)+p\frac{\cosh\left((\frac12-\alpha)\beta E\right)}{\cosh\left(\frac12\beta E\right)},
\end{equation}
where $p$ is defined in (\ref{def:p}). Its adjoint w.r.t. the duality $\bra A|B\ket=\tr AB$ is the completely positive operator on $\cB(\cH_\p)$ given by
$$
\cL_\alpha^\ast(B)=\tr_{\cH_\a}(I\otimes\rho_\beta^{1-\alpha})
\e^{\i\tau H}(B\otimes\rho_\beta^\alpha)\e^{-\i\tau H}.
$$
\eep

Going back to the special case $\alpha=0$, we have
$$
\rho_{\p,n\tau}=\cL_0^n(\rho_\p),
$$
so that the discrete semi-group $(\cL_0^n)_{n\in\N}$ acting on the density matrices of $\cH_\p$ 
describes the reduced Schr\"odinger dynamics of the particle. The reduced Heisenberg dynamics
is obtained by duality: for $B\in\cB(\cH_\p)$ and $\rho_\p\in\cB^1(\cH_\p)$,
$$
\tr_{\cH_\p} B\cL_0^n(\rho_\p)=\tr_{\cH_\p}\cL_0^{\ast n}(B)\rho_\p.
$$

At this point, the choice of $M$ becomes immaterial and we can consider an
arbitrary large number of interactions. Given an observable $B$ on $\cH_\p$, we write
$$
\langle B\rangle_n= \tr B \cL_0^n(\rho_\p), 
$$
for its expectation value at time $t=n\tau$. 


\section{Results and discussion}
\label{sec:results}

We are now in a position to state our main results on the dynamics of the particle.
As will be shown in Section \ref{s:oneatom}, coupling with a single
2-level atom turns the periodic Bloch oscillations (\ref{eq:xfreeevol}) of frequency 
$\omega_{\rm Bloch}=F$ into quasi-periodic motion with the two frequencies 
$\omega_{\rm Bloch}$ and
\begin{equation}
\omega_0=\sqrt{(E-F)^2+4\lambda^2},
\label{def:omega0}
\end{equation}
(see Equ. (\ref{eq:xcoupledevol}) below).
Repeated interactions with 2-level atoms have a much more drastic effect. The
bounded motion of the particle now becomes diffusive. In terms of the parameter
\begin{equation}
\label{def:p}
p=\frac{4\lambda^2}{\omega_0^2}\sin^2\left(\frac{\omega_0\tau}{2}\right)\in[0,1],
\end{equation}
the motion is characterized by a drift velocity
\begin{equation}
\vd=\vd(E,F)=\frac{p}{\tau} \tanh\left(\frac{\beta E}{2} \right),
\label{def:vd}
\end{equation}
and a diffusion constant
\begin{align}
 D&=D(E,F) =\frac{p}{2\tau}\left( 1-p\tanh^2\left(\frac{\beta E}{2}\right)\right).
\label{def:dd}
\end{align}
More precisely, the following holds.

\begin{theorem}\label{thm:main} 
Assume that $F>0$, $\lambda\not=0$ and $\omega_0\tau\notin2\pi\Z$ so that
$p\in]0,1]$. Let the density matrix $\rho_\p\in\cB^1(\cH_\p)$ describe the initial state of the particle
and denote by $\mu_n$ the spectral measures of the position observable $X$ in the state 
$\rho_{\p,n\tau}$,
\begin{equation}
\label{eq:mun}
 \mu_n(f) =\int f(x)\,\d\mu_n(x) = \langle f(X)\rangle_n.
\end{equation}
\begin{enumerate}[\rm 1.]
\item The Central Limit Theorem (CLT) holds: For any bounded continuous $f$ on $\R$,
\begin{equation*}
\lim_{n\to\infty}\int f\left(\frac{x-\vd n\tau}{\sqrt{2Dn\tau}}\right)\,\d\mu_n(x)
=\int f(x)\,\e^{-x^2/2}\,\frac{\d x}{\sqrt{2\pi}}.
\end{equation*}
\item If $\tr\left(X^2\rho_{\mathrm p}\right)<+\infty$, then
\begin{equation*}
\lim_{n\to\infty} \frac{\langle X\rangle_n}{n\tau}= \vd,\quad \lim_{n\to\infty} \frac{\bra (X-\vd n\tau)^2\ket_n}{n\tau}=2D.
\end{equation*}
\item If $\tr\left(\e^{\gamma |X|}\rho_\p\right)<+\infty$ for all $\gamma>0$ then a Large Deviation Principle
(LDP) holds in the sense that, for any interval $J\subset \R$,
\begin{equation}\label{eq:gellis}
\lim_{n\to\infty} \frac{1}{n}\log \mu_n(nJ)= - \inf_{x\in J} I(x),
\end{equation}
where $I(x)$ is the Legendre-Fenchel transform of $e(\eta)=\log\theta(-\eta/\beta E)$,
i.e., 
$$
I(x)=\sup_{\eta\in\R}\left[\eta x-\log\theta\left(-\frac{\eta}{\beta E}\right)\right].
$$
Here the function $\theta$ is defined in~\eqref{eq:thetadef}.
\end{enumerate}
\end{theorem}

\bigskip
Note that when $E=F$, the mobility
$$
\mu=\lim_{F\to 0} \frac{v_\d}{F}=\frac{\beta\sin^2(\lambda\tau)}{2\tau},
$$
and the diffusion constant
$$
D=\mu\beta^{-1}\left(1-\sin^2(\lambda\tau)\tanh^2\left(\frac{\beta F}{2}\right)\right),
$$
satisfy the Einstein relation
$$
\lim_{F\to 0} D=\mu\beta^{-1}=\mu k_B T.
$$
\medskip

The rate function in Part 3 is explicitly given by
$$
I(x)=\left\{
\begin{array}{ll}
\displaystyle -x\left(\frac{\beta E}{2}+\log\left(\frac{R(x)-x}{a(1-x)}\right)\right)
-\log\left(\frac{(1-p)(R(x)+1)}{1-x^2}\right)&\text{for } x\in[-1,1],\\[10pt]
+\infty&\text{otherwise},
\end{array}\right.
$$
where
$$
a=\frac{p}{(1-p)\cosh(\beta E/2)},\qquad R(x)=\sqrt{x^2+a^2(1-x^2)}.
$$
It is strictly convex on $[-1,1]$ and satisfies $I(\vd\tau)=0$ and $I(x)>0$ for $x\not= \vd\tau$.

Note that the drift velocity and diffusion constant do not depend on the initial state of the particle. 
The CLT gives us the probability to find the particle at time $n\tau$ in a region of size $O(\sqrt n)$ 
around the mean value $\vd n\tau$, whereas the LDP gives information on 
this probability for a region of size $O(n)$. To put it differently, it yields information 
on the probability that the particle's mean speed falls asymptotically in an interval of size $O(1)$.
Loosely speaking, it says that
$$
\mu_n(\{n(\vd+\delta v)\tau\})\simeq \e^{-nI((\vd+\delta v)\tau)}.
$$

The peculiar symmetry $e(-\beta E-\eta)=e(\eta)$ immediately leads to the relation 
$I(x)=-\beta Ex+ I(-x)$ which tells us that
$$
\lim_{\delta v\downarrow0}\lim_{n\to\infty}\frac1{n\tau}\log
\frac{\mu_n(n[-v-\delta v,-v+\delta v]\tau)}{\mu_n(n[v-\delta v,v+\delta v]\tau)}=-\beta E v,
$$
{\sl i.e.,} that negative mean velocities are exponentially less likely than positive ones.
The reader familiar with recent developments in non-equilibrium statistical mechanics
will recognize here a kind of fluctuation theorem. Indeed, we shall see that the symmetry
of the function $e(\eta)$ is a direct consequence of time-reversal invariance and, in a sense,
a remnant of the Evans-Searles (or transient) fluctuation theorem (see (\ref{eq:transient})).

We have further studied the statistics of the energy changes 
of the particle, the environment and the whole system. Note that the latter is not expected
to vanish, since the Hamiltonian is time-dependent, so that total energy is not conserved.

To study the change in the energy of the atomic reservoir we use the following operational procedure.
The reservoir being initially in thermal equilibrium at inverse temperature $\beta$ and the particle
in the state $\rho_\p$, we measure the total energy of the reservoir and the particle
just before the first interaction and just after the $n$-th interaction. These successive measurements 
yield the four values  $E_{\p,0},E_{\p,n}\in\sp\,H_\p$ and $E_{\env,0},E_{\env,n}\in\sp\,H_\env$.
It will be convenient to express the resulting change in energy in terms of the ``entropy like''
quantities
$$
\Delta S_{\p,n}=\beta^\ast(E_{\p,n}-E_{\p,0}),\qquad
\Delta S_{\env,n}=-\beta(E_{\env,n}-E_{\env,0}),
$$
where $\beta^\ast=\beta E/F$. We denote by $\P^n$ the joint probability distribution of 
$\Delta S_{\p,n}$, $\Delta S_{\env,n}$ and by $\E^n$ the corresponding expectation.
Note that the quantities $E_{\env,0}$ and $E_{\env,n}$ are well defined provided the reservoir contains
only a finite 
number $M$ of atoms and $n\leq M$. However, under those circumstances,  $\Delta S_{\env,n}$ does 
not depend on $M$ and we can therefore consider the reservoir contains an infinite number of atoms. This simple thermodynamical limit will always be
understood in what follows. 

{\noindent\bf Remark.} When applied to electric charge, or more generally to particle number,
the two measurement processes described above go under the name {\sl full counting
statistics} (see e.g. \cite{ABGK} and references therein). The present application is closer to
the approach to current fluctuations found in \cite{dR}.

\bet\label{thm:FCSenergy}
\begin{enumerate}[\rm 1.]
\item $\P^n[\Delta S_{\p,n}=\Delta S_{\env,n}]=1$. Hence, in the following we set
$$
\Delta S_n=\Delta S_{\p,n}=\Delta S_{\env,n}.
$$
\item The cumulant generating function of  $\Delta S_n$ is given by
$$
\log\E^n\left[\e^{\alpha\Delta S_n}\right]=n\log\theta(\alpha),
$$
where the function $\theta$ is defined in~\eqref{eq:thetadef}.
\item Its mean value and variance are
$$
\E^n\left[\frac{\Delta S_n}{n}\right]=-\beta E\vd\tau,\qquad
\E^n\left[\frac{(\Delta S_n+\beta E\vd n\tau)^2}{n}\right]
=(\beta E)^2 2D\tau.
$$
\item The CLT holds: For any bounded continuous function $f$,
$$
\lim_{n\to\infty}\E^n\left[
f\left(\frac{\Delta S_n+\beta E\vd n\tau}{\beta E\sqrt{2Dn\tau}}\right)\right]
=\int f(x)\,\e^{-x^2/2}\,\frac{\d x}{\sqrt{2\pi}}.
$$
\item The sequence $(\P^n)_{n\in\N}$ satisfies a LDP: For any intervals $J\subset\R$,
$$
\lim_{n\to\infty}\frac1n\log\P^n\left[\frac{\Delta S_n}{n}\in J\right]
=-\inf_{s\in J} \phi(s),
$$
with the rate function $\phi(s)=\sup_{\alpha\in\R}(\alpha s-\log\theta(\alpha))$.

\item It satisfies the transient fluctuation theorem
\begin{equation}\label{eq:transient}
\frac{\P^n\left[\frac{\Delta S_n}n=-s\right]}{\P^n\left[\frac{\Delta S_n}n=s\right]}=\e^{ns}.
\end{equation}
\end{enumerate}
\eet

Part 1 clearly reflects the fact that the number operator (\ref{def:Nop}) commutes with $H$
so that
$$
\beta^\ast H_\p+\beta H_\env=\beta E\left(
\frac{H_\p}{F}+\frac{H_\env}{E}\right),
$$
is preserved by the repeated interaction dynamics.

The particle's drift velocity is $v_{\mathrm d}$, and one sees therefore that, as expected, its energy
loss per unit time equals the work done by $F$ per unit time. Simultaneously, the environment gains
 energy at a rate $E\vd$: indeed, the particle moves on average $v_{\mathrm d}$ steps to the right
per unit time, which corresponds to $v_{\mathrm d}$ elements of the chain gaining an energy $E$. 
This leads to an average energy gain or loss of $(E-F)v_{\mathrm d}$ for the full system. In the 
special case $E=F$, these rates are equal, and the total system neither looses nor gains energy. 
This is a consequence of the fact (mentioned after Equ. (\ref{def:Nop})) that the interaction term 
in the Hamiltonian commutes with the free Hamiltonian in this case. In general, the total energy is not
 preserved, which is a reflection of the fact that the Hamiltonian of the total system is time-dependent, 
as is clear from~\eqref{eq:totalH}.

Note also that the symmetry $\theta(1-\alpha)=\theta(\alpha)$ which leads to the transient fluctuation theorem (\ref{eq:transient}) is evident from Proposition \ref{prop:Lalpha}. We shall see in Section \ref{ssec:tri} that it is actually a consequence of time-reversal invariance.

Since
$$
-\frac{H_\p}{F}\simeq X,
$$
we expect a very similar result for the position increment $\Delta X_n=X_n-X_0$
obtained from a double measurement of $X$ at time $t=0$ and $t=n\tau$.
Indeed, the distribution $\Q^n$ of $\Delta X_n$ satisfies the following.

\bet\label{thm:FCSx}
\begin{enumerate}[\rm 1.]
\item The cumulant generating function of  $\Delta X_n$ satisfies
$$
g(\eta)=\lim_{n\to\infty}\frac1n\log\Q^n\left[\e^{\eta\Delta X_n}\right]
=\log\theta\left(-\frac\eta{\beta E}\right).
$$
\item Its mean value and variance are
$$
\lim_{n\to\infty}\Q^n\left[\frac{\Delta X_n}{n\tau}\right]=\vd,\qquad
\Q^n\left[\frac{(\Delta X_n-\vd n\tau)^2}{n\tau}\right]=2D.
$$
\item The CLT holds: For any bounded continuous function $f$,
$$
\lim_{n\to\infty}\Q^n\left[
f\left(\frac{\Delta X_n-\vd n\tau}{\sqrt{2Dn\tau}}\right)\right]
=\int f(x)\,\e^{-x^2/2}\,\frac{\d x}{\sqrt{2\pi}}.
$$
\item The sequence $(\Q^n)_{n\in\N}$ satisfies a LDP: For any intervals $J\subset\R$,
$$
\lim_{n\to\infty}\frac1n\log\Q^n\left[\frac{\Delta X_n}{n}\in J\right]
=-\inf_{x\in J} I(x).
$$

\item It satisfies the asymptotic fluctuation theorem
$$
\beta E(v-\delta v)\le
\lim_{n\to\infty}
\frac1n\log\frac{\Q^n\left[\frac{\Delta X_n}{n\tau}\in[-v-\delta v,-v+\delta v]\right]}
{\Q^n\left[\frac{\Delta X_n}{n\tau}\in[v-\delta v,v+\delta v]\right]}
\le\beta E(v+\delta v),
$$
for $v\in]-1,1[$ and $\delta v>0$, small enough.
\end{enumerate}
\eet

{\noindent\bf Remark.} This fluctuation theorem is not of transient (or Evans-Searles) type since it only holds in the large time limit. It is not of the stationary (or Gallavotti-Cohen) type either since there is no stationary state for the dynamics, as we shall see in Section \ref{ssec:kraus}. For more details on these fluctuation theorems, we refer the reader to e.g. \cite{JPR,RM}.

Note also the similarity with Theorem \ref{thm:main}. Theorem \ref{thm:FCSx} describes the position increment of the particle without requiring sharp localization of the particle position at time $0$, contrary to Theorem \ref{thm:main}.

\section{Interaction with a single two level atom}
\label{s:oneatom}

In this section we investigate the dynamics of the particle interacting with a single atom
described by the Hamiltonian (\ref{eq:honeatom}). As already remarked, $H$ can easily be
diagonalized by exploiting the fact that it commutes with the number operator (\ref{def:Nop}).
To get a tractable formula for the propagator $\e^{\i tH}$, it will be more convenient
to consider the unitary operator
$$
U=(Tb^\ast b+bb^\ast)\cos\theta-(Tb^\ast-b)\sin\theta,
$$
where $\theta$ is chosen such that
$$
\cos(2\theta)=\frac{E-F}{\omega_0},
\qquad
\sin(2\theta)=\frac{2\lambda}{\omega_0}.
$$
Using the commutation relations (\ref{PartCR}) and (\ref{CAR}), one easily shows that
\begin{equation}
\label{eq:hdiag}
U^*HU=H_\p +\omega_0\left(b^\ast b-\frac12\right)+\frac{E-F}{2}.
\end{equation}
It follows immediately that, for $F\not=0$, the spectrum of $H$ is purely discrete,
\begin{equation*}
\sp(H)=2-F\Z +\frac{E-F}{2} \pm \frac{\omega_0}{2}.
\end{equation*}
If $\omega_0/F$ is not an integer, all the eigenvalues are simple. The normalized
eigenvectors are given by
\begin{align*}
|\phi_{k,-}\ket &= U|\psi_k\ket\otimes\gs=
\cos\theta |\psi_k\ket\otimes \gs-\sin\theta |\psi_{k+1}\ket\otimes\es,\\
|\phi_{k,+}\ket &= U|\psi_k\ket\otimes\es=
\cos\theta |\psi_{k+1}\ket\otimes\es+\sin\theta |\psi_k\ket\otimes\gs,
\end{align*}
where $\gs$ and $\es$ denote the ground state and the excited state of the atom.
Of course the situation is completely different when $F=0$. The spectrum
$$
\sp(H)=\left(\frac12(E-F-\omega_0)+[0,4]\right)\bigcup
\left(\frac12(E-F+\omega_0)+[0,4]\right),
$$
is then purely absolutely continuous.
\medskip

  From Equ. (\ref{eq:hdiag}) we get the explicit formula for the propagator,
\begin{equation}
\label{eq:unitarygroup}
\e^{\i tH}=U\e^{\i t(E-F)/2}\e^{\i t\omega_0(b^\ast b-1/2)}\e^{\i tH_\p}U^\ast,
\end{equation}
which reduces the proof of the following result to a straightforward calculation.

\begin{lemma}\label{lem:unitarygroup}
For any operator $A$ on $\mathcal H_\p$ and any $t\in\R$ one has
$$
\e^{-\i tH}A\otimes \rho_\beta\ \e^{\i tH}=\cA_{\beta,t}(A_t)\otimes b^*b
+\cB_{\beta,t}(A_t)\otimes b
+\cB_{\beta,t}(A_t^\ast)^*\otimes b^*+\cC_{\beta,t}(A_t)\otimes bb^*,
$$
where $A_t=\e^{-\i tH_\p}A\,\e^{\i tH_\p}$ and
\begin{eqnarray*}
\cA_{\beta,t}(A) & = & \frac{\e^{-\beta E}}{Z_\beta} \left(
1-\frac{4\lambda^2}{\omega_0^2}\sin^2\left(\frac{\omega_0 t}{2}\right)\right)A
+\frac{1}{Z_\beta}\frac{4\lambda^2}{\omega_0^2}\sin^2\left(\frac{\omega_0 t}{2}\right)TAT^*,\\
\cB_{\beta,t}(A) & = & \frac{1}{Z_\beta}\frac{2\lambda}{\omega_0}\left(
\frac{\i}2\sin(\omega_0 t)
-\frac{E-F}{\omega_0}\sin^2\left(\frac{\omega_0 t}{2}\right)\right)
\left(AT^*-\e^{-\beta E}T^*A\right),\\
\cC_{\beta,t}(A) & = & \frac{1}{Z_\beta}\left(
1-\frac{4\lambda^2}{\omega_0^2}\sin^2\left(\frac{\omega_0 t}{2}\right)\right)A
+\frac{\e^{-\beta E}}{Z_\beta}\frac{4\lambda^2}{\omega_0^2}
\sin^2\left(\frac{\omega_0 t}{2}\right) T^\ast AT.
\end{eqnarray*}
\end{lemma}

By noting that $X=2X\otimes\rho_{\beta=0}$ the preceding lemma can be used to
compute the evolution of the position observable
\begin{align}
\e^{\i tH}X\,\e^{-\i tH}&=\e^{\i tH_\p}X\e^{-\i tH_\p}\nonumber\\
&+\left(\frac{4\lambda^2}{\omega_0^2}(bb^\ast-b^\ast b)
+\frac{2\lambda(E-F)}{\omega_0^2}(Tb^\ast+T^\ast b)\right)\sin^2\left(\frac{\omega_0t}{2}\right)
\label{eq:xcoupledevol}\\
&-\i\frac{\lambda}{\omega_0}(Tb^\ast-T^\ast b)\sin(\omega_0t).\nonumber
\end{align}
We conclude that the coupling to a single atom does not substantially alter the long term 
behavior of the particle. In particular, when $F\not=0$, the motion remains bounded. 
We will see in the next section that the situation is very different for repeated interactions
with a sequence of atoms.


\section{Repeated interaction dynamics}
\label{s:proofs}

In this section we study the properties of the operator $\cL_\alpha$, defined in~\eqref{def:Lalpha}. 

It is clear that $\alpha\mapsto\cL_\alpha$ is entire analytic as a map from $\C$ to the
bounded operators on $\cB(\cH_\p)$. Moreover, since
$$
\tr_{\cH_\p}B\cL_\alpha(A)
=\tr(B\otimes\rho_\beta^{\alpha})\e^{-\i\tau H}(A\otimes\rho_\beta^{1-\alpha})\,\e^{\i\tau H},
$$
$\cL_\alpha$ is also bounded as an operator on the Schatten-von Neumann class $\cB^p(\cH_\p)$,
for any $p\in[1,\infty]$.

Using the cyclicity of the partial trace w.r.t. operators 
on $\cH_\a$ we can write, for $\alpha\in\R$,
$$
\cL_\alpha(A)=
\tr_{\cH_\a}\left(\e^{-\i\tau H^{(\alpha)}}A\otimes\rho_\beta\,\e^{\i\tau H^{(\alpha)\ast}}\right),
$$
where $H^{(\alpha)}$ is a bounded perturbation of $H$,
\begin{align*}
H^{(\alpha)}&=\e^{-\alpha\beta H_\a/2}H\e^{\alpha\beta H_\a/2}\\
&=H_\p+H_\a+\lambda\e^{-\alpha\beta H_\a/2}V\e^{\alpha\beta H_\a/2}\\
&=H_\p+H_\a+\lambda(\e^{-\alpha\beta E/2}Tb^\ast+\e^{\alpha\beta E/2}T^\ast b).
\end{align*}
This shows in particular that $\cL_\alpha$ is completely positive for real values of $\alpha$.

\subsection{Gauge invariance}

\bel\label{lem:Lcommut} The operator $\cL_\alpha$ commutes with the evolution of the
non-interacting particle, i.e.,
$$
\cL_\alpha\left(\e^{-\i tH_\p}A\,\e^{\i tH_\p}\right)=\e^{-\i tH_\p}\cL_\alpha(A)\e^{\i tH_\p},
$$
holds for all $t,\alpha\in\R$ and $A\in\cB(\cH_\p)$.
\eel

{\noindent\bf Proof.} From the fact that $H$ commutes with the number operator (\ref{def:Nop})
we infer
$$
H=\e^{\i tN}H\e^{-\i tN}=\e^{\i tH_\p/F}\e^{\i tH_\a/E}H\e^{-\i tH_\a/E}\e^{-\i tH_\p/F},
$$
so that
$$
\e^{-\i tH_\a/E}\e^{-\i\tau H}\e^{\i tH_\a/E}=\e^{\i tH_\p/F}\e^{-\i\tau H}\e^{-\i tH_\p/F}.
$$
Since $H_\p$ and $H_\a$ commute, we also have
$$
\e^{-\i tH_\a/E}\e^{-\i\tau H^{(\alpha)}}\e^{\i tH_\a/E}
=\e^{\i tH_\p/F}\e^{-\i\tau H^{(\alpha)}}\e^{-\i tH_\p/F},
$$
and hence
\begin{align*}
\e^{\i tH_\p}\cL_\alpha&\left(\e^{-\i tH_\p}A\,\e^{\i tH_\p}\right)\e^{-\i tH_\p}\\
&=\tr_{\cH_\a}\left(\e^{\i tH_\p}\e^{-\i\tau H^{(\alpha)}}\e^{-\i tH_\p}A
\otimes\rho_\beta\,\e^{\i tH_\p}\e^{\i\tau H^{(\alpha)\ast}}\e^{-\i tH_\p}\right)\\
&=\tr_{\cH_\a}\left(\e^{-\i tH_\a F/E}\e^{-\i\tau H^{(\alpha)}}\e^{\i tH_\a F/E}A
\otimes\rho_\beta\,\e^{-\i tH_\a F/E}\e^{\i\tau H^{(\alpha)\ast}}\e^{\i tH_\a F/E}\right)\\
&=\tr_{\cH_\a}\left(\e^{-\i tH_\a F/E}\e^{-\i\tau H^{(\alpha)}}A\otimes\rho_\beta\,
\e^{\i\tau H^{(\alpha)\ast}}\e^{\i tH_\a F/E}\right)\\
&=\cL_\alpha(A),
\end{align*}
where we used the fact that $H_\a$ commutes with $A\otimes\rho_\beta$ and the cyclicity 
of the trace.
\qed

Introducing the non-interacting evolution operator
$$
\cU(A)=\e^{-\i\tau H_\p}A\,\e^{\i\tau H_\p},
$$
we define the ``interaction picture'' reduced evolution as
$\widetilde\cL_\alpha=\cL_\alpha\circ\cU^{-1}$.
Note that, by Lemma \ref{lem:Lcommut}, we have
$$
\cL_\alpha^n=\widetilde\cL_\alpha^n\circ\cU^n=\cU^n\circ\widetilde\cL_\alpha^n,
$$
for any $n\in\N$.

\subsection{Time-reversal invariance}\label{ssec:tri}

Let us denote by $C_\p$ the complex conjugation on $\ell^2(\Z)$, {\sl i.e.,} 
$(C_\p\psi)(x)=\overline{\psi(x)}$ and set
$$
\cC(A)=C_\p AC_\p^\ast,
$$
for $A\in\cB(\cH_\p)$. This anti-linear involution implements time reversal of the particle's dynamics. 
Indeed, since the Hamiltonian $H_\p$ is real in the position representation, $C_\p H_\p=H_\p C_\p$, 
one has $C_\p\e^{\i tH_\p}=\e^{-\i tH_\p}C_\p$ for all $t\in\R$ and in particular
$$
\cC\circ\cU=\cU^{-1}\circ\cC.
$$

\bel\label{lem:TRI}
For all $\alpha\in\R$ one has $\cL^\ast_\alpha=\cC\circ\cL_{1-\alpha}\circ\cC$.
\eel

{\bf\noindent Remark.} As will be clear from its proof, this property of $\cL_\alpha$ is a consequence 
of the time-reversal invariance of the dynamics of the particle coupled to a two-level atom.
It implies that the spectral radius of $\cL_\alpha$ and $\cL_{1-\alpha}$ coincide.

{\bf\noindent Proof.} Setting $C_\a(a_0|0\ket+a_1|1\ket)=\overline{a_0}|0\ket+\overline{a_1}|1\ket$
defines a anti-unitary operator on $\cH_\a$ such that $C_\a H_\a=H_\a C_\a$ and hence
$$
C_\a\e^{\i tH_\a}=\e^{-\i tH_\a}C_\a,\qquad C_\a\rho_\beta=\rho_\beta C_\a.
$$
Since $T$ and $T^\ast$ are real w.r.t. $C_\p$ and $b$ and $b^\ast$ are real w.r.t. $C_\a$,
the Hamiltonian $H$ is real w.r.t. $C=C_\p\otimes C_\a$ and one has $C \e^{\i tH}=\e^{-\i tH}C$.

From the fact that the partial trace on $\cH_\a$ satisfies
$$
\tr_{\cH_\a}C AC^\ast=\cC(\tr_{\cH_\a}A),
$$
for all $A\in\cB(\cH_\p\otimes\cH_\a)$, one deduces that, for $A,B\in\cB(\cH_\p)$,
\begin{align*}
\cC(A\cL_\alpha(B))
&=\cC(\tr_{\cH_\a}(A\otimes\rho_\beta^{\alpha})
\e^{-\i\tau H}(B\otimes\rho_\beta^{1-\alpha})\e^{\i\tau H})\\
&=\tr_{\cH_\a}C(A\otimes\rho_\beta^{\alpha})
\e^{-\i\tau H}(B\otimes\rho_\beta^{1-\alpha})\e^{\i\tau H}C^\ast\\
&=\tr_{\cH_\a}(\cC(A)\otimes\rho_\beta^{\alpha})
\e^{\i\tau H}(\cC(B)\otimes\rho_\beta^{1-\alpha})\e^{-\i\tau H}\\
&=\tr_{\cH_\a}(\cC(B)\otimes\rho_\beta^{1-\alpha})\e^{-\i\tau H}
(\cC(A)\otimes\rho_\beta^{\alpha})\e^{\i\tau H}\\
&=\cC(B)\cL_{1-\alpha}\circ\cC(A).
\end{align*}
It follows that
\begin{align*}
\tr_{\cH_\p }A\cL_\alpha(B)
&=\overline{\tr_{\cH_\p }\cC(A\cL_\alpha(B))}\\
&=\overline{\tr_{\cH_\p }\cC(B)\cL_{1-\alpha}\circ\cC(A)}\\
&=\tr_{\cH_\p }\cC\circ\cL_{1-\alpha}\circ\cC(A)B,
\end{align*}
and hence $\cL_{\alpha}^\ast(A)=\cC\circ\cL_{1-\alpha}\circ\cC(A)$.
\qed


\subsection{Kraus representation}\label{ssec:kraus}

Since $\cL_\alpha$ is completely positive for $\alpha\in\R$, the same is true of $\widetilde\cL_\alpha$.
The following result. describes the Kraus representation of the latter operator.

\begin{lemma} \label{lem:reduceddyn} 
For any $\alpha\in\R$ and $A\in\cB(\mathcal H_\p)$, one has 
\begin{equation}
\label{eq:reduceddyn}
\widetilde\cL_\alpha(A)=\e^{\alpha\beta E}p_-T^*A T + p_0 A + \e^{-\alpha\beta E}p_
+ TA T^*,
\end{equation}
where
\begin{equation*}
p_-=\frac{\e^{-\beta E}}{1+\e^{-\beta E}}\,p,\quad p_0=1-p, \quad p_+=\frac{1}{1+\e^{-\beta E}}\,p.
\end{equation*}
with $p$ defined in Equ.~\eqref{def:p}.
\end{lemma}

We note that $\e^{\alpha\beta E}p_-+p_0+\e^{-\alpha\beta E}p_+=\theta(\alpha)$, where $\theta(\alpha)$ is defined in Proposition \ref{prop:Lalpha}, so
\begin{equation}
\label{equ:LstaronI}
\cL_\alpha(I)=\cL_\alpha^\ast(I)=\theta(\alpha)I.
\end{equation}
Consider $\cK_\alpha=\theta(\alpha)^{-1}\cL_\alpha$ as an operator on $\cB^1(\cH_\p)$.
Since $\cK_\alpha$ is completely positive and trace preserving, it has unit spectral radius
(see \cite{Sch}). Hence, $\theta(\alpha)$ is the spectral radius of $\cL_\alpha$. This proves Proposition \ref{prop:Lalpha}. The remark after Lemma \ref{lem:TRI} now explains the origin of the symmetry $\theta(1-\alpha)=\theta(\alpha)$.

Let us now set $\alpha=0$ in Equ. (\ref{eq:reduceddyn}) and explore the implications of this
expression for the dynamics of the particle. If $\rho$ describes the state of the particle, then 
$T^*\rho T$ (respectively $T\rho T^*$) represent the same state translated by one lattice spacing
to the left (respectively right). Note moreover that 
$$
p_-+p_0+p_+=1,
$$
so that the reduced evolution $\cL_0=\widetilde\cL_0\circ\cU$ consists of a free evolution with the 
Hamiltonian $H_\p$, followed by a random translation by $\pm 1$ or $0$, and with probabilities 
$p_{\pm}$ or $p_0$.  Note that the dynamics is trivial if $p=0$, {\sl i.e.,} if $\omega_0\tau=2\pi m$
with $m\in\Z$. In that case there is no translation and the particle evolves according to $H_\p$.
This can be seen directly on Equ. (\ref{eq:unitarygroup}) by noticing that
$UH_\p U^\ast=H_p+Fb^\ast b$. It follows that the propagator factorizes
$$
\e^{\i\tau H}=(-1)^m \e^{\i\tau(E-F)/2} \, \e^{\i\tau H_\p} \otimes\e^{\i\tau Fb^*b},
$$
and, up to an inessential phase factor and a renormalization of the atomic Bohr frequency, the particle
and the two-level system evolve as if they were not coupled. This resembles the ``Rabi oscillation'' 
phenomenon which appears in the Jaynes-Cummings model for matter-radiation interaction.
In the following we will avoid this resonance and assume $p\not=0$. 

Introducing a family of i.i.d. random variables $Y_n$, taking the values $\pm 1$ and $0$ with 
probability $p_\pm$ and $p_0$, and defining $S_n=\sum_{j=1}^n Y_j$, we can very concisely write 
\begin{equation}
\label{eq:semigroup}
\cL_0^n(\rho) = \e^{-\i n\tau H_\p}\E\left[T^{S_n}\rho\,T^{-S_n}\right]\e^{\i n\tau H_\p}. 
\end{equation}
Accordingly, the study of the dynamics of the system is reduced to that of a classical random walk. 
As a further remark, suppose that the initial state of the particle is invariant under the uncoupled 
dynamics, so that
\begin{equation}\label{eq:invstate}
\rho=\sum_{k\in\Z} p_k |\psi_k\rangle\langle \psi_k|.
\end{equation}
Then, using Equ. (\ref{eq:Tpsi}), we obtain 
$\cL_0^n(\rho)=\sum_{k\in\Z} p_{k}^{(n)}|\psi_k\rangle\langle \psi_k|$ with
\begin{equation}
\label{eq:master}
p_{k}^{(n+1)}=p_+p_{k-1}^{(n)}+p_0p_{k}^{(n)}+p_-p_{k+1}^{(n)}.
\end{equation}
Thus, the set of $H_\p$-invariant states is invariant under the reduced dynamics and the latter
reduces to a classical Markov chain on this set.

Before turning to the proof of Lemma \ref{lem:reduceddyn}, let us show that the reduced dynamics 
has no stationary state: there exists no density matrix $\rho$ on $\cH_\p$ such that $\cL_0(\rho)=\rho$.
Indeed, it follows from Lemma \ref{lem:Lcommut} that the subspaces $\cJ_d$, $d\in\Z$, defined by
\begin{align*}
\cJ_d = & \{\rho \in{\cal B}^1(\cH_\p)\,|\,
\e^{-\i tH_\p}\rho\,\e^{\i tH_\p}=\e^{\i td}\rho\text{ for all }t\in\R\} \\
= & \{\rho \in{\cal B}^1(\cH_\p)\,|\, \rho=\sum_{k\in\Z} \rho_k |\psi_k \ket\bra \psi_{k+d}| \},
\end{align*}
are globally invariant under $\cL_0$. Hence, if a state $\rho$ is stationary, so is its diagonal part 
$\rho_0=\sum_k p_{k} |\psi_k\ket\bra \psi_k|$, where $p_k=\bra\psi_k|\rho\psi_k\ket$. 
Equ. (\ref{eq:master}) then writes
$$
p_{k-1}-Z_\beta p_k+\e^{-\beta E} p_{k+1}=0,
$$
which implies that $p_k=a+b\e^{\beta Ek}$ for some constants $a,b\in\R$. But this contradicts
the fact that $1=\tr\,\rho=\sum_kp_k$.

\bigskip
\noindent{\bf Proof of Lemma~\ref{lem:reduceddyn}.}
We start with the fact that
$$
\rho_\beta^{\alpha}\e^{-\i\tau H}A\otimes\rho_\beta^{1-\alpha}\,\e^{\i\tau H}=
\frac{Z_{\alpha\beta}Z_{(1-\alpha)\beta}}{Z_\beta}\rho_{\alpha\beta}\,
\e^{-\i\tau H}A\otimes\rho_{(1-\alpha)\beta}\,\e^{\i\tau H},
$$
so that, applying Lemma \ref{lem:unitarygroup}, we get
\begin{align*}
\rho_\beta^{\alpha}\e^{-\i\tau H}A\otimes\rho_\beta^{1-\alpha}\,\e^{\i\tau H}=
\frac{Z_{\alpha\beta}Z_{(1-\alpha)\beta}}{Z_\beta}\rho_{\alpha\beta}\bigl(
&\cA_{(1-\alpha)\beta,t}(A_t)b^\ast b
+\cB_{(1-\alpha)\beta,t}(A_t)b\\
+&\cB_{(1-\alpha)\beta,t}(A_t)^\ast b^\ast
+\cC_{(1-\alpha)\beta,t}(A_t)bb^\ast
\bigr).
\end{align*}
Upon taking the partial trace over $\cH_\a$, we obtain
$$
\cL_\alpha(A)=\frac{Z_{\alpha\beta}Z_{(1-\alpha)\beta}}{Z_\beta}
\left(
\cA_{(1-\alpha)\beta,t}(A_t)\frac{\e^{-\alpha\beta E}}{Z_{\alpha\beta}}
+\cC_{(1-\alpha)\beta,t}(A_t)\frac{1}{Z_{\alpha\beta}}
\right),
$$
and the result follows from Lemma \ref{lem:unitarygroup} with some elementary algebra.
\qed

\section{Proof of Theorem~\ref{thm:main}}\label{s:proofthm1}

To complete the proof of our main results, we shall need the following technical lemma.

\bel\label{lem:expo}1. If $F\not=0$ then, for any $\eta\in\R$, one has
$$
\lim_{t\to\infty}\|\e^{\i tH_\p}\e^{\i\eta X/\sqrt{t}}\e^{-\i tH_\p}-\e^{\i\eta X/\sqrt{t}}\|=0.
$$
2. If $F\not=0$ then, for any $\eta\in\R$, there exists $C_\eta\ge1$ such that
$$
C_\eta^{-1}\le\e^{-\eta X/2} \e^{\i tH_\p}\e^{\eta X}\e^{-\i tH_\p}\e^{-\eta X/2}\le C_\eta,
$$
for all $t\in\R$.
\eel

{\noindent\bf Proof.} 1. By Equ. (\ref{eq:xfreeevol}), one has
$\e^{\i tH_\p}X\e^{-\i tH_\p}=X+B_t$, where $B_t$ is a uniformly bounded
operator valued function of $t$. Duhamel formula yields
$$
\e^{\i tH_\p}\e^{\i\eta X/\sqrt{t}}\e^{-\i tH_\p}
-\e^{\i\eta X/\sqrt{t}}=\frac{\eta}{\sqrt t}R_t,
$$
where
\begin{align*}
R_t=\int_0^1&\e^{\i(1-s)\eta\left(X+B_t\right)/\sqrt{t}}B_t\,\e^{\i s\eta X/\sqrt{t}} \d s.
\end{align*}
The claim follows from the fact that $R_t$ is also uniformly bounded.

2. By Equ. (\ref{HpConjug}) we can write, for any $\eta,t\in\R$,
$$
Q(t,\eta)=\e^{\i tH_\p}\e^{-\i\eta X}\e^{-\i tH_\p}\e^{\i\eta X}=
\e^{\i G}\e^{-\i tFX}\e^{-\i G}\e^{-\i\eta X}\e^{\i G}\e^{\i tFX}\e^{-\i G}\e^{\i\eta X}.
$$
From the commutation relation (\ref{PartCR}) we get, for $\theta\in\R$,
$$
\e^{-\i\theta X}G\,\e^{\i\theta X}=\frac2F\sin(\xi+\theta),
$$
so that
$$
Q(t,\eta)=\e^{2\i(\sin\xi-\sin(\xi+tF)-\sin(\xi+\eta)+\sin(\xi+tF+\eta))/F}.
$$
It follows that $\eta\mapsto Q(t,\eta)\in\cB(\cH_\p)$ extends to an entire analytic function. 
Moreover, one easily shows that
$$
C_\eta=\sup_{t\in\R}\|Q(t,\eta)\|=\e^{4\sinh(|\mathrm{Im}\eta|)/F},
$$
for any $\eta\in\C$.

Now since $X_t=\e^{\i tH_\p}X\e^{-\i tH_\p}$ is self-adjoint the subspace 
$\cD_t=\mathrm{Dom}(\e^{X_t^2})$ is dense in $\cH_\p$ and such that, for $\phi\in\cD_t$,
the vector valued function $\eta\mapsto\e^{\i\eta X_t}\phi$ is entire analytic.
For $\psi\in\cD_t$, $\phi\in\cD_0$ and $\eta\in\R$ one has
$$
\bra\e^{\i\overline{\eta}X_t}\psi|\e^{\i\eta X}\phi\ket=\bra\psi|Q(t,\eta)\phi\ket.
$$
By analytic continuation, both sides of this identity extend to complex values of $\eta$.
In particular, one has
$$
\bra\e^{\eta X_t}\psi|\e^{-\eta X}\phi\ket=\bra\psi|Q(t,\i\eta)\phi\ket,
$$
for any $\eta\in\R$. Since $\cD_t$ is a core of $\e^{\pm\eta X_t}$,
the last identity extends to all $\phi\in\mathrm{Dom}(\e^{-\eta X})$ and
$\psi\in\mathrm{Dom}(\e^{\eta X_t})$. The modulus of its
right hand side being bounded by $C_{\i\eta}\|\psi\|\,\|\phi\|$, one concludes that
$\mathrm{Ran}(\e^{-\eta X})\subset\mathrm{Dom}(\e^{\eta X_t})$ and
$$
\|\e^{\eta X_t}\e^{-\eta X}\|\le C_{\i\eta}.
$$
It follows that
$$
\e^{-\eta X/2}\e^{\eta X_t}\e^{-\eta X/2}
=(\e^{\eta X_t/2}\e^{-\eta X/2})^\ast(\e^{\eta X_t/2}\e^{-\eta X/2})\le C_{\i\eta/2}^2,
$$
and
$$
(\e^{-\eta X/2}\e^{\eta X_t}\e^{-\eta X/2})^{-1}=\e^{\eta X/2}\e^{-\eta X_t}\e^{\eta X/2}
\le C_{-\i\eta/2}^2=C_{\i\eta/2}^2,
$$
which together imply
$$
C_{\i\eta/2}^{-2}\le\e^{-\eta X/2}\e^{\eta X_t}\e^{-\eta X/2}\le C_{\i\eta/2}^2.
$$
\qed

\bigskip
We now turn to the proof of Theorem \ref{thm:main}.
First, from Equ. (\ref{eq:semigroup}) we find that
$$
\bra f(X)\ket_{n}= \tr (f(X)\cL_0^n(\rho_{\mathrm p}))=
\E\left[ \tr \left( f(T^{-S_n}\e^{in\tau H_{\mathrm p}} X \e^{-in\tau H_{\mathrm p}} T^{S_n})
 \, \rho_\p  \right) \right].
$$
Hence, using Equ. (\ref{eq:xfreeevol}), we get
\begin{align}
\bra f(X)\ket_{n}
&=\E\left[\tr \left(f\left(T^{-S_n}\left(X+B_n\right)T^{S_n}\right)\rho_\p\right)\right]\nonumber\\
&=\E\left[\tr\left(f\left(X+S_n+B_n\right)\rho_\p\right)\right],
\label{eq:expectvalue}
\end{align}
where
\begin{equation}\label{eq:Bn}
B_n=\frac{4}{F}\sin\left(\frac{n\tau F}{2}\right)
\sin\left(\xi+\frac{n\tau F}{2}\right).
\end{equation}
To prove Part 1, we study the convergence in distribution of the sequence of probability measures 
$\tilde\mu_n(J)=\mu_n(\sqrt{2Dn\tau}J+\vd n\tau)$, $J\subset \R$, where $\mu_n$ is defined 
in~\eqref{eq:mun}. By the L\'evy-Cram\'er continuity theorem, 
this is equivalent to the pointwise convergence of the sequence of their 
characteristic functions. We shall therefore prove that for any $\eta\in\R$
\begin{equation*}
\tilde\mu_n(\e^{\i\eta x})=\int_\R\e^{\i\eta x}\,\d\tilde\mu_n(x) 
=\left\bra\e^{\i\eta \left(X-\vd n\tau\right)/\sqrt{2Dn\tau}}\right\ket_{n}  
\stackrel{n\to\infty}{\longrightarrow} \e^{-\eta^2/2}.
\end{equation*}
Using (\ref{eq:expectvalue}), we can write
\begin{align}
\tilde\mu_n(\e^{\i\eta x})
&=\E\left[\tr\left(\e^{\i\eta\left(X+S_n-\vd n\tau+B_n\right)/\sqrt{2Dn\tau}}\rho_\p\right)\right]\nonumber\\
&=\E\left[\e^{\i\eta(S_n-\vd n\tau)/\sqrt{2Dn\tau}}\right]
\tr\left(\e^{\i\eta\left(X+B_n\right)/\sqrt{2Dn\tau}}\rho_\p\right).
\label{eq:characteristic2}
\end{align}
The classical CLT implies that the first factor of the last line converges, as $n\to\infty$, to 
$\e^{-\eta^2/2}$ since, as a simple computation confirms
\begin{equation}
\label{eq:SnStats}
\frac{1}{n\tau}\E[S_n]=\vd,\qquad
\frac{1}{n\tau} {\rm Var}[S_n]=2 D,
\end{equation}
where $\vd$ and $D$ are defined in~\eqref{def:vd} and~\eqref{def:dd}. The second 
factor is controlled by Part 1 of Lemma \ref{lem:expo} which implies
$$
\lim_{n\to\infty}\tr\left(\e^{\i\eta\left(X+B_n\right)/\sqrt{2Dn\tau}}\rho_\p\right)
=\lim_{n\to\infty}\tr\left(\e^{\i\eta X/\sqrt{2Dn\tau}}\rho_\p\right).
$$
The right hand side of this identity is $1$ by the dominated convergence theorem.

\bigskip
The first assertion in Part  2 follows from
\begin{align*}
\frac1{n\tau}\bra X\ket_n&=
\frac1{n\tau}\tr(X\cL_0^n(\rho_\p))\\
&=\frac1{n\tau}\E\left[\tr((X+S_n+B_n)\rho_\p)\right]\\
&=\frac1{n\tau}\E[S_n]+\frac1{n\tau}\tr((X+B_n)\rho_\p),
\end{align*}
since the first term on the last line converges to $v_d$ (by Equ. (\ref{eq:SnStats})) while the 
second vanishes as $n\to\infty$. The proof of the second assertion is similar.

\bigskip
To prove Part~3, we shall study the cumulant generating function
\begin{equation}
\label{def:lmgf}
e_n(\eta)=\log\mu_n(\e^{\eta x})=\log\tr\left(\e^{\eta X}\cL_0^n(\rho_\p)\right).
\end{equation}
By Equ. (\ref{eq:expectvalue})-(\ref{eq:Bn}) and~\eqref{eq:xfreeevol}, one has
$$
e_n(\eta)=\log\E\left[\tr\left(\e^{\eta\left(X+S_n+B_n\right)}\rho_\p\right)\right]
 =\log\E\left[\e^{\eta S_n}\right]+\log\tr\left(\e^{\i n\tau H_\p}\e^{\eta X}\e^{-\i n\tau H_\p}\rho_\p\right).
$$
One easily computes the first term on the right hand side 
$$
\log\E\left[\e^{\eta S_n}\right]=n\log\E\left[\e^{\eta Y_1}\right]
=n\log\theta\left(-\frac{\eta}{\beta E}\right).
$$
Writing the second term as
$$
\log\tr\left((\rho_\p^{1/2}\e^{\eta X/2})\,
\e^{-\eta X/2}\e^{\i n\tau H_\p}\e^{\eta X}\e^{-\i n\tau H_\p}\e^{-\eta X/2}\,
(\e^{\eta X/2} \rho_\p^{1/2})\right),
$$
and applying Part 2 of Lemma \ref{lem:expo} we get the bound
$$
|\log\tr\left(\e^{\i n\tau H_\p}\e^{\eta X}\e^{-\i n\tau H_\p}\rho_\p\right)|\le \log C_\eta
+\log\tr\left(\e^{\eta X}\rho_\p\right),
$$
from which we conclude that
$$
e(\eta)=\lim_{n\to\infty}\frac1n e_n(\eta)
=\log\theta\left(-\frac{\eta}{\beta E}\right).
$$
Since $e(\eta)$ is differentiable, the G\"artner-Ellis theorem (see e.g. \cite{DZ}) 
implies the LDP \eqref{eq:gellis} with the rate function $I(x)$ related to $e(\eta)$ via the 
Legendre-Fenchel transform. Finally, the symmetry $\theta(1-\alpha)=\theta(\alpha)$
translates into
$$
e(\eta)=e(-\beta E-\eta),
$$
which implies $I(x)=-\beta Ex +I(-x)$.

\section{Full counting statistics}
\label{s:FCS}

In this section we start with a precise description of the two measurements processes involved
in the formulation of Theorems \ref{thm:FCSenergy} and \ref{thm:FCSx} 
and proceed then to the proofs of these results.

Suppose that the initial state of the particle is described by the density matrix $\rho_\p$ and set 
$\rho=\rho_\p\otimes\rho_\beta^{\otimes M}$. Let $A_1,\ldots,A_m$ be commuting
self-adjoint operators on the Hilbert space $\cH=\cH_\p\otimes\cH_\env$. We assume the $A_j$
to have pure point spectrum.  We define the vector valued observable $A=(A_1,\ldots,A_m)$
and its spectrum $\sp\,A=\sp\,A_1\times\cdots\times\sp\,A_m$. We denote by $P_a$ the spectral
projection associated to the eigenvalue $a\in\sp\,A$.

The outcome of a first measurement of the observables $A$ at time $t=0$ will  be $a\in\sp\,A$
with probability $\tr\,\rho P_a$. After this measurement the state of the combined system is 
reduced to $\rho'=P_a\rho P_a/\tr(\rho P_a)$. This state now evolves under the repeated interaction 
dynamics and, after the $n$-th interaction, becomes $U(n\tau,0)\rho'U(n\tau,0)^\ast$.
A second measurement of $A$ at time $t=n\tau$ will yield the result $a'$ with
probability $\tr\,U(n\tau,0)\rho'U(n\tau,0)^\ast P_{a'}$. Thus, the joint probability
distribution for the two successive measurements of $A$ is given by
\begin{align*}
\P_A^n(a,a')&=\tr\,\rho P_a\times\tr\,U(n\tau,0)\rho'U(n\tau,0)^\ast P_{a'}\\[5pt]
&=\tr\,U(n\tau,0)P_a\rho P_a U(n\tau,0)^\ast P_{a'}.
\end{align*}
Therefore, the probability distribution of the measured increment $\Delta a=a'-a$ after 
$n$ interactions is
$$
\P_A^n(\Delta a)=\sum_{a,a'\in\sp\,A\atop a'-a=\Delta a}
\tr\,U(n\tau,0)P_a\rho P_a U(n\tau,0)^\ast P_{a'}.
$$
The cumulant generating function of this distribution is 
\begin{align*}
g_n(\alpha)&=\log \sum_{\Delta a\in\sp\,A-\sp\,A}
\P_A^n(\Delta a)\,\e^{\alpha\cdot\Delta a}\\[5pt]
&=\log\sum_{a,a'\in\sp\,A}\e^{\alpha\cdot(a'-a)}\,\tr\,U(n\tau,0)P_a\rho P_a U(n\tau,0)^\ast P_{a'},
\end{align*}
where $\,\cdot\,$ denotes the Euclidean scalar product on $\R^m$.
At this point, it is useful to note that
$$
\widetilde\rho=\sum_{a\in\sp\,A}P_a\rho P_a,
$$
is a density matrix which commutes with $A$ so that $P_a\rho P_a=\widetilde\rho P_a$.
Hence, we can rewrite
\begin{align}
g_n(\alpha)&=\log\sum_{a,a'\in\sp\,A}\e^{\alpha\cdot(a'-a)}\,
\tr\,U(n\tau,0)\widetilde\rho P_a U(n\tau,0)^\ast P_{a'}\nonumber\\
&=\log\tr\,U(n\tau,0)\widetilde\rho\,\e^{-\alpha\cdot A}
U(n\tau,0)^\ast\e^{\alpha\cdot A}.\label{equ:cumulgen}
\end{align}
In the special case where the $A_j$ are observables of the particle, we obtain
\begin{equation}
\label{equ:cumulpart}
g_n(\alpha)=\log\tr\,\cL_0^n(\widetilde\rho_\p\,\e^{-\alpha\cdot A})\e^{\alpha\cdot A}
=\log\tr\,\widetilde\rho_\p\,\e^{-\alpha\cdot A}\cL_0^{\ast n}(\e^{\alpha\cdot A}),
\end{equation}
where $\widetilde\rho_\p=\sum_{a\in\sp\,A}P_a\rho_\p P_a$.

\subsection{Proof of Theorem \ref{thm:FCSenergy}}

To prove theorem \ref{thm:FCSenergy} we consider the case
$A=(\beta^\ast H_\p,-\beta H_\env)$. From Equ. (\ref{equ:cumulgen}), and 
for any $n\leq M$ (where $M$ is the number of atoms in the reservoir),  we have
\begin{align*}
g_n(\alpha_\p,\alpha_\env)&=\log\E^n\left[\e^{\alpha_\p\Delta S_{\p,n}
+\alpha_\env\Delta S_{\env,n}}\right]\\
&=\log\tr\,U(n\tau,0)\widetilde\rho\,\e^{-\alpha_\p\beta^\ast H_\p+\alpha_\env\beta H_\env}
U(n\tau,0)^\ast\e^{\alpha_\p\beta^\ast H_\p-\alpha_\env\beta H_\env},
\end{align*}
where
$$
\widetilde\rho=\sum_{E\in\sp\,H_\p\atop E'\in\sp\,H_\env}
P_{\p,E}\rho_\p P_{\p,E}\otimes P_{\env,E'}\rho_\beta^{\otimes M}P_{\env,E'},
$$
and $P_{\p,E}$, $P_{\env,E'}$ are the spectral projections of $H_\p$ and $H_\env$.
Since $\rho_\beta^{\otimes M}$ commutes with $H_\env$, this reduces to
$\widetilde\rho=\widetilde\rho_\p\otimes\rho_\beta^{\otimes M}$ with
$\widetilde\rho_\p=\sum_{E\in\sp\,H_\p}P_{\p,E}\rho_\p P_{\p,E}$. Hence,
invoking (\ref{LalphaMeans}), we obtain
\begin{align*}
&g_n(\alpha_\p,\alpha_\env)\\
&=\log\tr\,U(n\tau,0)(\widetilde\rho_\p\,\e^{-\alpha_\p\beta^\ast H_\p}
\otimes\rho_\beta^{\otimes M}\e^{\alpha_\env\beta H_\env})
U(n\tau,0)^\ast(\e^{\alpha_\p\beta^\ast H_\p}\otimes\e^{-\alpha_\env\beta H_\env})\\
&=\log\tr\,(\e^{\alpha_\p\beta^\ast H_\p}\otimes\left[\rho_\beta^{\alpha_\env}\right]^{\otimes M})
U(n\tau,0)(\widetilde\rho_\p\e^{-\alpha_\p\beta^\ast H_\p}
\otimes\left[\rho_\beta^{1-\alpha_\env}\right]^{\otimes M})U(n\tau,0)^\ast\\
&=\log\tr_{\cH_\p}\, \e^{\alpha_\p\beta^\ast H_\p}\cL_{\alpha_\env}^n
(\widetilde\rho_\p\,\e^{-\alpha_\p\beta^\ast H_\p})\\
&=\log\tr_{\cH_\p}\, \cU^{\ast n}(\e^{\alpha_\p\beta^\ast H_\p})
\widetilde\cL_{\alpha_\env}^n(\widetilde\rho_\p\,\e^{-\alpha_\p\beta^\ast H_\p})\\
&=\log\tr_{\cH_\p}\, \e^{\alpha_\p\beta^\ast H_\p/2}
\widetilde\cL_{\alpha_\env}^n(\e^{-\alpha_\p\beta^\ast H_\p/2}\widetilde\rho_\p\,
\e^{-\alpha_\p\beta^\ast H_\p/2})\e^{\alpha_\p\beta^\ast H_\p/2}.
\end{align*}
Again, the number $M$ of interactions is now irrelevant and we may consider arbitrary values of $n$. The commutation relations (\ref{PartCR}) imply $\e^{\eta H_\p}T\e^{-\eta H_\p}=\e^{-\eta F}T$
so that, by (\ref{eq:reduceddyn}),
\begin{align*}
\e^{\alpha_\p\beta^\ast H_\p/2}\widetilde\cL_{\alpha_\env}&(\e^{-\alpha_\p\beta^\ast H_\p/2}A\,
\e^{-\alpha_\p\beta^\ast H_\p/2})\e^{\alpha_\p\beta^\ast H_\p/2}\\
&=\e^{(\alpha_\p+\alpha_\env)\beta E}p_-T^\ast AT+p_0A
+\e^{-(\alpha_\p+\alpha_\env)\beta E}TAT^\ast\\
&=\widetilde\cL_{\alpha_\p+\alpha_\env}(A).
\end{align*}
It follows that
$$
g_n(\alpha_\p,\alpha_\env)
=\log\tr\, I\widetilde\cL_{\alpha_\p+\alpha_\env}^n(\widetilde\rho_\p)
=\log\tr\,\widetilde\rho_\p\widetilde\cL_{\alpha_\p+\alpha_\env}^{\ast n}(I)
=n\log\theta(\alpha_\p+\alpha_\env),
$$
from which we conclude that
$$
\log\E^n\left[\e^{\alpha(\Delta S_{\p,n}-\Delta S_{\env,n})}\right]=n\log\theta(0)=0,
$$
which proves Part 1, from which Part 2,
$$
g_n(\alpha)=\log\E^n\left[\e^{\alpha\Delta S_{\p,n}}\right]=n\log\theta(\alpha),
$$
immediately follows.
Differentiation of the last identity at $\alpha=0$ gives Part 3.
Since $\theta(\alpha)$ is an entire function of $\alpha$ such that $\theta(0)=1$,
$\log\theta(\alpha)$ is an analytic function of $\alpha$ in a complex 
neighborhood of $0$ and Part 4 follows from the Bryc theorem (see \cite{Bry}).
The G\"artner-Ellis theorem directly applies to give Part 5.
Finally, Part 6 is a direct consequence of the symmetry $\theta(1-\alpha)=\theta(\alpha)$
which implies
\begin{align*}
\log\sum_{s}\P^n\left[\Delta S_{\p,n}=-s\right]\e^{-s}\e^{\alpha s}
&=\log\sum_{s}\P^n\left[\Delta S_{\p,n}=s\right]\e^{(1-\alpha)s}\\
&=g_n(1-\alpha)=g_n(\alpha)=\log\sum_s\P^n\left[\Delta S_{\p,n}=s\right]\e^{\alpha s},
\end{align*}
and hence $\P^n\left[\Delta S_{\p,n}=-s\right]\e^{-s}=\P^n\left[\Delta S_{\p,n}=s\right]$.

\subsection{Proof of Theorem \ref{thm:FCSx}}

We now consider the case where $A=X$. The cumulant generating function of the 
increment $\Delta X_n$ is given by Equ. (\ref{equ:cumulpart}),
$$
g_n(\eta)=\log\Q^n\left[\e^{\eta\Delta X_n}\right]
=\log\tr\,\cL_0^n(\widetilde\rho_\p\e^{-\eta X})\e^{\eta X}.
$$
Using the factorization $\cL_0^{n}=\cU^{n}\circ\widetilde\cL_0^{n}$, we further get
\begin{align*}
g_n(\eta)
&=\log\tr\,\widetilde\cL_0^n(\e^{-\eta X/2}\widetilde\rho_\p\,\e^{-\eta X/2})\cU^{\ast n}(\e^{\eta X})\\
&=\log\tr(\e^{\eta X/2}\widetilde\cL_0^n(\e^{-\eta X/2}\widetilde\rho_\p\,\e^{-\eta X/2})\e^{\eta X/2})
(\e^{-\eta X/2}\e^{\i n\tau H_\p}\e^{\eta X}\e^{-\i n\tau H_\p}\e^{-\eta X/2}).
\end{align*}
We note that $\e^{\eta X}T\e^{-\eta X}=\e^{\eta}T$ and 
$\e^{\eta X}T^\ast\e^{-\eta X}=\e^{-\eta}T^\ast$ so that, by (\ref{eq:reduceddyn}),
$$
\e^{\eta X/2}\widetilde\cL_0(\e^{-\eta X/2}A\,\e^{-\eta X/2})\e^{\eta X/2}
=\e^{-\eta}p_-T^\ast AT+p_0A+\e^{\eta}TAT^\ast
=\widetilde\cL_{-\eta/\beta E}(A).
$$
It follows that
$$
g_n(\eta)=\log\tr\,\widetilde\cL_{-\eta/\beta E}^n(\widetilde\rho_\p)
(\e^{-\eta X/2}\e^{\i n\tau H_\p}\e^{\eta X}\e^{-\i n\tau H_\p}\e^{-\eta X/2}).
$$
Part 2 of Lemma \ref{lem:expo} yields the estimates
$$
\log C_{\eta}^{-1}\,\tr\,\widetilde\cL_{-\eta/\beta E}^n(\widetilde\rho_\p) I
\le g_n(\eta)\le\log C_\eta\,\tr\,\widetilde\cL_{-\eta/\beta E}^n(\widetilde\rho_\p) I,
$$
and since $\tr\,\widetilde\cL_{-\eta/\beta E}^n(\widetilde\rho_\p) I
=\tr\,\widetilde\rho_\p\widetilde\cL_{-\eta/\beta E}^{\ast n}(I)=\theta(-\eta/\beta E)^n$,
we finally get
$$
\frac1n g_n(\eta)=\log\theta\left(-\frac\eta{\beta E}\right)+O\left(\frac1n\right),
$$
so that
$$
g(\eta)=\lim_{n\to\infty}\frac1n g_n(\eta)=\log\theta\left(-\frac\eta{\beta E}\right),
$$
which proves Part 1. Part 2 and Part 4 follow from the G\"artner-Ellis theorem while
Part 3 follows from the Bryc theorem. Finally, the LDP implies
$$
\lim_{n\to\infty}
\frac1n\log\frac{\Q^n\left[\frac{\Delta X_n}{n}\in[-q-\delta,-q+\delta]\right]}
{\Q^n\left[\frac{\Delta X_n}{n}\in[q-\delta,q+\delta]\right]}
=-\inf_{|x-q|\le\delta}I(-x)+\inf_{|x-q|\le\delta}I(x)
$$
and the fact that $I(-x)=I(x)+\beta Ex$ leads to Part 5.

\end{document}